\documentstyle[12pt,epsf]{article}

\def\be{\begin{eqnarray}}
\def\ee{\end{eqnarray}}

\def\q{{\bf q}}
\def\He#1{$^#1$He}

\begin{document}

\title{Representation of the short-range interactions
in liquid helium via modified hard sphere potentials}

\author{I.~O.~Vakarchuk, A.~A.~Rovenchak$^*$ \\
Department for Theoretical Physics,
Ivan Franko Lviv National University,\\
12 Draghomanov Str., Lviv, Ukraine, UA--79005,
tel.: +380 322 979443.
}

\maketitle

\begin{abstract}
In this paper we propose five different modifications of the hard
sphere potential for the modeling a short-range repulsion
and the calculation of thermodynamic and
transport properties of liquid \He4. We calculate the potential
energy, the total energy, and the sound velocity at $T=0$~K. It is
shown that three of the proposed potentials give a satisfactory
description of these properties.

\end{abstract}


\section{Introduction}
In this paper we calculate some properties of the helium system
using the modifications of helium--helium interatomic potential
obtained in~\cite{JPS2000}.

The method of receiving the potential is based on the collective variables
formalism as described in our previous work~\cite{JPS2000}.

Numerical methods could be grouped in two large parts: 1)~calculations
with adjustable parameters using experimental data on
thermodynamic and transport properties~\cite{Aziz79,Aziz91,Aziz92,Aziz97}, 2)~{\it ab
initio} quantum-mechanical computations (mainly Monte Carlo) \cite{Anderson93,Tang95}.
A good comparison between various potentials, calculated
properties of helium, and the experimental data might be found
in~\cite{Boronat94}.

We analyze our potentials due to the form of the potential curve
and such properties as the potential energy, the total energy, and
the sound velocity, all calculated at $T=0$~K, i.~e., for the ground state.

Concerning the form of the potential curve, one can notice that
the short-range part of the potential is not very significant
starting from distances about 2~\AA{} and less. This statement becomes clear if
one looks at the pair distribution function $F_2(R)$ of
helium~\cite{F2exp}: $F_2(R<2.212{\rm~\AA})=0$.

On the other hand, the results for the potential
from~\cite{JPS2000} lead to the value $-7.8$~K for the total energy of the
system of helium atoms at 0~K, while the experiment gives
$-7.17$~K~\cite{ExpData}, we consider it as a good agreement.
But we obtained the value of the potential energy
$-15.5$~K, and computer simulations~\cite{Boronat94} predict it in the range from
about $-20$ to $-22$~K (there is no experimental measurements available for this
quantity). This difference may not seem to be quite satisfactory
if one calculates quantities connected with the potential energy
only.

Thus, we tried to correct our potential for it could reflect the
short-range interactions more precisely unlike that obtained
in~\cite{JPS2000}. For this purpose we used some modification of the
hard sphere potential.

The paper is organized as follows. The calculating procedure is
described in the Section~2. Different forms for the short-range
repulsion are presented in the Section~3, the respective
potential curves are also given there. In the Section~4 we adduce
the computation of the energy and the sound velocity of \He4 in the
ground state.

\section{Calculating procedure}

The following relation was used for the calculation of the potential
Fourier image $\nu_q$. see~\cite{JPS2000,TMF2}:
\be\label{FirstNu}
\frac{N}{V}\frac{m\nu_q}{\hbar^2}&+&{\bf q}^2\,a_2({\bf q})-
{\bf q}^2\,a_2^2({\bf q})-\frac{1}{2N}\sum_{\bf k \neq 0}
        {\bf k}^2\,a_4({\bf k, -k, q, -q})  \\ \nonumber
&&
{}+\frac{1}{N}\sum_{\bf k \neq 0}{\bf k(-k-q)}\,a_3({\bf q, k, -k-q})=0,
\ee
$N$ is the total number of particles, $V$ is the volume of the
system, in the thermodynamic limit we have $N/V={\rm
const}=\rho=0.02185$~\AA$^{-3}$; $m$ is the mass of \He4 atom;
the quantities $a_3, a_4$ are expressed through $a_2$, see~\cite{JPS2000,TMF2}.

We are going to solve this problem in the approximation of ``two sums over the
wave vector". Such calculations were made in~\cite{JPS2000} with the
expression for $a_2$ in the form of
\be
a_2({\bf q})={1\over 2}\left(1-{1\over S_q}\right)-{1\over 2}\Sigma({\bf q}),
\ee
$\Sigma({\bf q})$ being complex functional of $a_2$ containing summation
(integration) over the wave vector~\cite{JPS2000,TMF2}.

The equation (\ref{FirstNu}) for $\nu_q$ is solved using iterative procedure.
We have used the corrected expression
for the zeroth approximation of the quantity $a_2$~\cite{TMFS0}:
\be
a^0_2(\q)={1\over 2}\left({1\over S_q^{0}}-{1\over S_q}\right),
\ee
while in~\cite{JPS2000} we used
\be
a^0_2(\q)={1\over 2}\left(1-{1\over S_q}\right);
\ee
$S_q^{0}$ stands for the short-range structure factor, i.~e.
it corresponds to the short-range repulsive interactions. The question
obtaining this structure factor $S_q^{0}$ is elucidated in the next section.

\section{Short-range potential}

\subsection{Hard sphere repulsion}

In this paper we first consider a short-range potential $\Phi^0(R)$ in the form of hard
sphere interaction,
\be \label{PhiHS}
\Phi^0(R)=\left\{
\begin{array}{cl}
\infty, & R\leq a\\
 0, & R > a
\end{array}\right. ,
\ee
$a$ is the hard sphere diameter.

This leads to $S_q^{0} = S_q^{\rm HS}$, the structure factor of
the
hard sphere system. It was calculated from the pair distribution
function of the hard sphere system in which we neglected all
terms with orders of density higher than 2. This approximation
will be called here as $\rho^2$-approximation or HS--$\rho^2$.

It is known that the Percus--Yevick approximation (PY)
gives quite good description in the case of short-range
interactions. It has also another advantage: there is an exact
solution of the Ornstein--Zernike equation for the hard sphere system in
PY~\cite{Balescu}. We denote the structure factor obtained in this
fashion as $S_q^{\rm PY}$.

The quantities $S_q^{\rm HS}-1$ and $S_q^{\rm PY}-1$ show a weak
damping $\propto 1/q^2$ at the large values of the wave vector.
It leads to the complication of the calculation procedure since
one should extend the integration limits in the Fourier
transformation of $\nu_q$. More precisely speaking, the damping of
$\nu_q$ was weak too and we used the extrapolation at large $q$ to
obtain correct behaviour of $\Phi(R)$ at small distances.

\begin{figure}[h]
\epsfysize=50mm
\centerline{\epsfbox{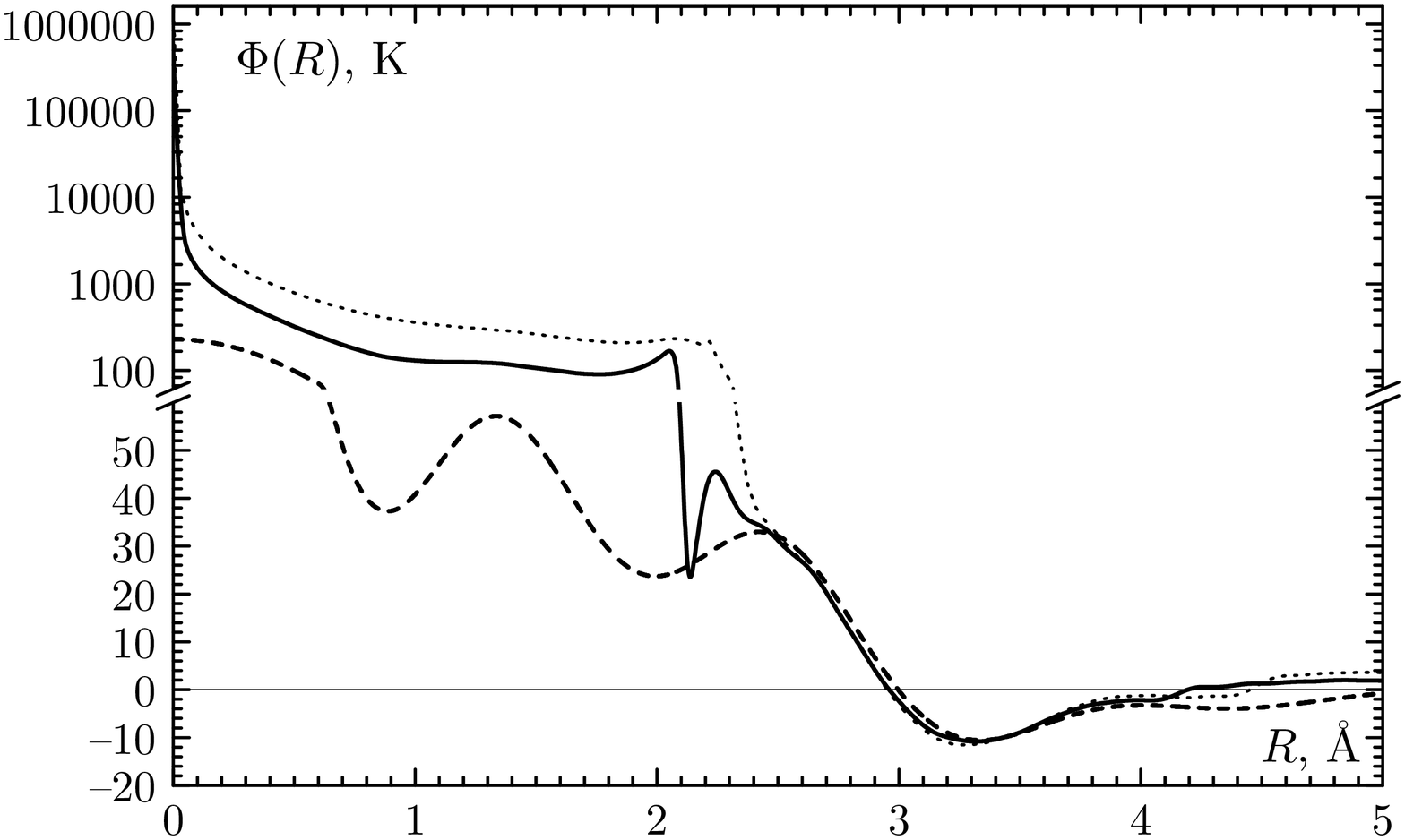}\qquad\epsfysize=50mm\epsfbox{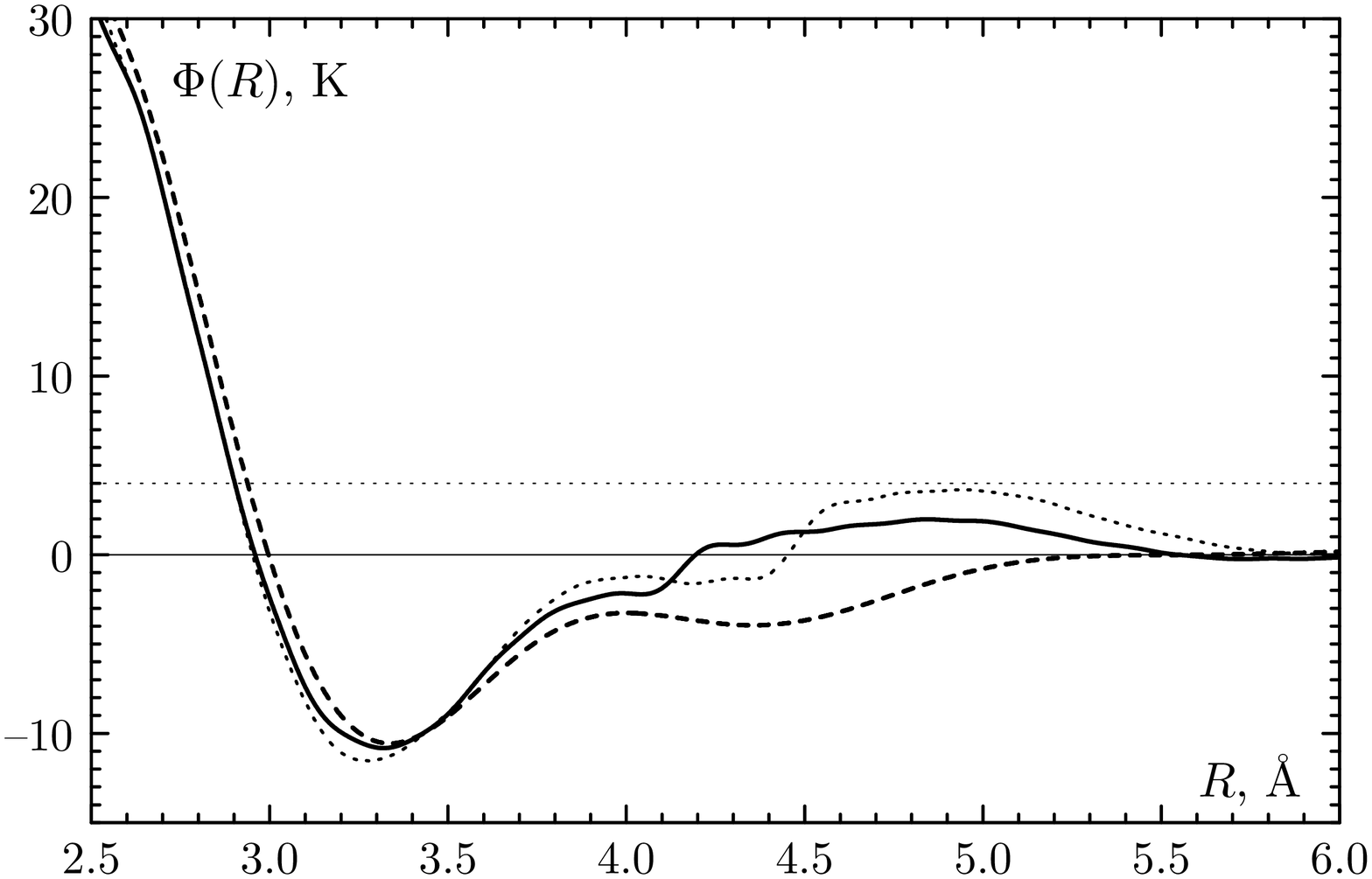}}
\centerline{a) \hspace{10cm} b)}
\caption{The potential $\Phi(R)$ with short-range interactions in hard sphere approximation
(HS-$\rho^2$ --- solid line, PY --- dotted line)
compared with results of our previous work~\protect\cite{JPS2000} (dashed line).}
\label{PhiHSPY}
\end{figure}

We have calculated the Fourier image $\nu_q$ and the potential $\Phi(R)$ using
the hard sphere structure factor both in HS--$\rho^2$ and PY approximations.
As one can see from the Fig.~\ref{PhiHSPY}, the
long-range behaviour of this potential is not satisfactory unlike the short-range one.

One can see from Fig.~\ref{PhiHSPY} the extra hill on the potential curve at approximately
6~\AA. We will try to find out the reasons of its appearing.

\bigskip
\bigskip

\subsection{Softened repulsion}

The true repulsive part of the potential should be a bit
``softened''. It is easier to formulate this ``softening" in the
terms of Meyer function $f(R)$:
\be
f(R)=e^{-\beta\Phi(R)}-1
\ee
that stays finite while $\Phi(R)$ can have infinities, see
(\ref{PhiHS}).

For the hard sphere system $f(R)$ reads
\be \label{fHS}
f^{\rm HS}(R)=\theta(R-a)-1=\left\{
\begin{array}{cl}
-1, & R\leq a\\
 0, & R > a
\end{array}\right. ,
\ee

In this terms, the softening means that the straight step of the Meyer
function of hard sphere must be slightly ``rounded". Here we
consider the variants of this ``rounding".

Let us suppose the power dependence in $\Phi^0(R)$: $\Phi^0(R)\propto
1/R^n$. Then the kinetic energy operator in the Schr\"odinger equation
gives:
\be
-{\hbar^2\over2m}{\partial^2\over\partial R^2}\,e^{-(A/R)^n}=
-{\hbar^2\over2m}{n^2\over R^2}\left({A\over R}\right)^{2n}e^{-(A/R)^n} +
O(1/R^{n+2}).
\ee
Thus, the power $2n+2$ must equal the power of inverse distance in
the repulsive part of real potential. Assuming it to be the
Lennard--Jones potential we obtain $n=5$.

Another way is to round the corners in the Meyer function via
direct accepting its form. For instance, like in the previous
case:
\be \label{fpow}
f(R)=e^{-(A/R)^n}-1.
\ee
The value of $n$ should be taken large (larger than 5) in order to resemble the
hard sphere step.

The first case (Lennard--Jones repulsion, $n=5$) will be called ``soft
spheres'' (SS), and the second one (directly accepted repulsion)
will be denoted as ``almost hard spheres'' (AHS).

\begin{figure}[h]
\epsfxsize=75mm
\centerline{\epsfbox{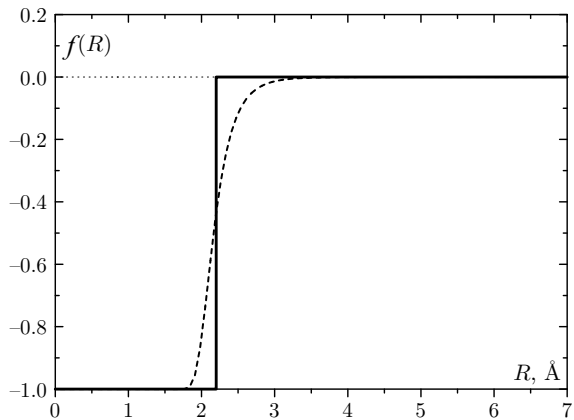}}
\caption{Meyer functions of hard spheres (solid line) and almost hard spheres (dashed line).}
\label{fAHS}
\end{figure}

Let us choose the following values for AHS quantities in (\ref{fpow}):
\be
A=2.1\ {\rm\AA},\qquad n=12.
\ee
The Meyer function then becomes quite close to the hard sphere
step, see Fig.~\ref{fAHS}. That is the reason why we call it ``almost hard
spheres''.

\bigskip
\bigskip

\begin{figure}[h]
\epsfxsize=75mm
\centerline{\epsfbox{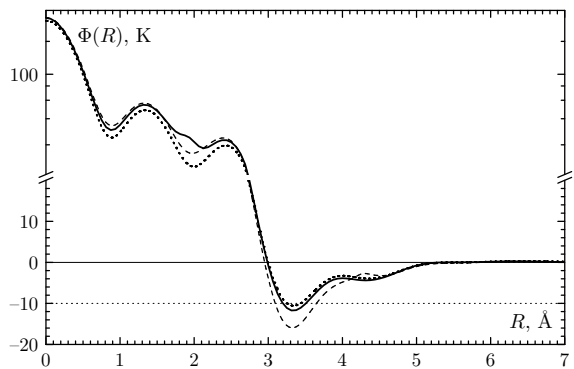}}
\caption{The potential $\Phi(R)$ with short-range interactions in
SS (dashed line) and AHS (solid line) approximation
compared with results of our previous work~\protect\cite{JPS2000} (dotted line)}
\label{PhiSSAHS}
\end{figure}

We have calculated the structure factor of the softened repulsive part using the PY
approximation. Since we do not need big accuracy, only the first
iteration in the Ornstein--Zernike equation was used.
Our results show that the potential curve does not differ much
from that of our previous work \cite{JPS2000}, see
Fig.~\ref{PhiSSAHS}.

\bigskip
\bigskip

In spite of large difference in $1/R$ powers (5 and 12), the
results do not yield to much different behaviour in the region of
small $R$: it still resembles that of \cite{JPS2000}. From this point
of view, the reason must be in principal difference between soft
and hard spheres, namely in the analyticity or non-analyticity of
Meyer function that leads to different types of damping at large $q$ for
the structure factor and, as a result, for the Fourier image
$\nu_q$.

Another fact concerns the region of potential well. One can see
that the increasing of $1/R$ power from 5 to 12 moves the depth of
the well closer to that of obtained from the hard sphere
approximation. On the other hand, similar value was obtained in
\cite{JPS2000} and it appears not to differ much from the results
of computer simulations \cite{Anderson93,Tang95} and the
computations with adjustable parameters
\cite{Aziz79,Aziz91,Aziz92,Aziz97}. Those results might be found in Table~1.

\bigskip
\noindent
\begin{center}
\begin{tabular}{|l|c|c|c|}
  \hline\hline
  \vphantom{\large\AA}%
  Potential & $\Phi(0)$, K  & Well              & Extra Hill$$       \\
  \hline
  \vphantom{\large\AA}%
  Ref.~\ref{JPS2000}&$\simeq240$& $-10.59$~K at 3.34~\AA & $0.26$~K at 6.41~\AA \\
  HS -- $\rho^2$&$\infty$   & $-10.86$~K at 3.32~\AA & $1.99$~K at 4.84~\AA \\
  HS -- PY      &$\infty$   & $-11.55$~K at 3.27~\AA & $3.64$~K at 4.94~\AA \\
  AHS           &$\simeq240$& $-11.76$~K at 3.34~\AA & $0.23$~K at 6.31~\AA \\
  SS            &$\simeq230$& $-15.95$~K at 3.32~\AA & $0.27$~K at 6.44~\AA \\
  Glued         &$\infty$   & $-11.72$~K at 3.33~\AA & $0.23$~K at 6.31~\AA \\
  \hline
  \vphantom{\large\AA}%
  Ref.~\ref{Anderson93}  & ---   & $-11.01$~K at 3.0\phantom{0}~\AA & no hill    \\
  Ref.~\ref{Aziz97}  &$2\cdot10^6$&$-11.04$~K at 2.99~\AA& no hill           \\
  \hline\hline
\end{tabular}

\medskip
Table~1. Parameters of the potential curve for different types of
potential.
\end{center}
\bigskip

The extra-hill was obtained on our potential curves at the
interatomic distances of $\simeq5-7$~\AA. We
consider it as a consequence of solving the problem of interaction
between helium atoms via quantum equations. But the large values of
hill maxima obtained in the case of hard-sphere interaction
should be treated as the result of Meyer function non-analyticity.
We claim it since the correspondent values in the case of soft spheres
and almost hard spheres are commensurable with the value from
\cite{JPS2000}.

The region of small $R$ values is probably the most complicated
one. It seems that one should accept the hard-sphere approximation
in order to obtain a good description here. We tried to join the advantages of the PY short-range part with
those of the AHS long-range part by just ``gluing'' this two data sets
at the point of 3.25~\AA. The left side corresponds to PY; the right side -- to AHS.

One can notice that there is a constant positive difference
between the minimum
position (the equilibrium interatomic distance) of the potentials discussed
in this work and of those calculated by Aziz {\it et al}. We consider it
as a consequence of the effective nature of our potentials. Let us consider
the model of hard spheres with diameter $a$ in order to illustrate this statement.
The equilibrium distance between two atoms in the system of two atoms is
obviously $a$. But if one considers the system of three particles than
the average distance slightly increases, namely by 5 per cent. Thus, the
difference in approximately 10 per cent in our case might be at least
partially considered
as a consequence of such a geometric effect.

\section{Energy and sound velocity}

One can calculate the potential energy $\Phi$ of the system at 0~K using the
relation~\cite{JPS1996}:
\be
\Phi=N{\rho\nu_0\over2}+\sum_{\bf q\neq0}
{\rho\nu_q\over2}\left({1\over\alpha_q}-1\right),\\
\alpha_q=\sqrt{1+{2\rho\nu_q \over \varepsilon_q}},
  \qquad
  \varepsilon_q={\hbar^2q^2\over 2m}.
\ee

The total energy $E$ reads:
\be
E=\Phi+ \sum_{\bf q\neq0} {\varepsilon_q\over 4}\,{(\alpha_q-1)^2\over\alpha_q}.
\ee

The quantity $\rho\nu_0$ in these
equation must be substituted with $mc^2$ where $c$ is the sound
velocity in \He4. The value $mc^2=\rho\nu_0$ corresponds to the
RPA and we want to have the next, postRPA approximation. Thus, the
corrected expression reads~\cite{TMFS0}:
\be
mc^2=\rho\nu_0-{1\over 8N}\sum_{\bf q\neq0}
\varepsilon_q{(\alpha_q-1)^2\over\alpha_q}.
\ee

We give the results discussed above in the Table~1 for the
comparison. The left hand values of sound velocity correspond to the
RPA ($mc^2=\rho\nu_0$),
while the right-hand values reflect the correction~\cite{TMFS0}.

\bigskip
\noindent
\begin{center}
\begin{tabular}{|l|c|r|c|}
  \hline\hline
  \vphantom{\large\AA}%
  Potential & $\Phi/N$, K&$E/N$, K&$c$, m/s\\
  \hline
  \vphantom{\large\AA}%
  Ref.~\ref{JPS2000}&$-15.5$ &$-7.8\ $  &292 vs 238\\
  HS-$\rho^2$ $^{\rm a})$   &$+18.1$ &$+38.8\ $ &407 vs 524\\
  AHS           &$-21.6$ &$-12.0\ $ &297 vs 240\\
  SS            &$-29.1$ &$-18.6\ $ &254 vs 182\\
  Glued         &$-16.7$ &$-8.0\ $  &324 vs 262\\
  \hline
  \vphantom{\large\AA}%
  Ref.~\ref{Boronat94}$\,^{\rm b})$  &$-21.7$  & $-7.1\ $  &241\\
  \hline
  \vphantom{\large\AA}%
  Experiment    &         &$-7.17\,^{\rm c})$&240.3$\,^{\rm d})$\\
  \hline\hline
\end{tabular}
\end{center}
\medskip
$^{\rm a}$){\footnotesize We do not give data for the potential obtained using $S^{\rm PY}$
since in this case one cannot obtain a ``good'' $\alpha_q$: value
of $1+2\rho\nu_q/\varepsilon_q$ is negative for some values of
$q$.}\\
$^{\rm b})${\footnotesize Calculation using potential~\cite{Aziz79}}.\\
$^{\rm c})${\footnotesize Ref.~\ref{ExpData}}.\\
$^{\rm d})${\footnotesize This value corresponds to the sound velocity
at saturated vapour pressure for $T=0.1$~K~\cite{McCarthy}.
Ref.~\ref{NIST} gives 237.8~m/s at $T=0.8$~K. These values are not strictly experimental but
it is possible to consider them as the averaged experimental ones since they are calculated using
the well-analyzed equation of state.}
\medskip

\begin{center}
Table~2. Energy and sound velocity with different potentials.
\end{center}
\bigskip

As might be seen from the Table~2, the HS model for the
short-range repulsion yields to incorrect values for energy and
the sound velocity. Unfortunately, we did not find the universal
potential satisfying the requirements to describe
the total energy and the potential energy with similar accuracy.
The best fit for both these quantities is found in the case of
``glued'' potential but the method of its obtaining looks very
artificially. One can see that the best fit of the
potential energy is found in the AHS approximation for the
short-range repulsion. On the other hand, previously obtained
potential~\cite{JPS2000} remains the best of those considered here
due to the values of the total energy and the sound velocity.

Therefore, three potential will be used in our further
works: the potential obtained in~\cite{JPS2000}, the
potential with AHS short-range repulsion, and the ``glued''
potential. The comprehensive analysis of thermodynamic and transport properties
that might be calculated and compared with the experiment or the computer simulations
will judge the most suitable model potential.


\begin{thebibliography}{99}
\item[$^*$]e-mail: {\tt andrij@ktf.franko.lviv.ua}\,.

\bibitem{JPS2000}I.~O.~Vakarchuk, V.~V.~Babin, A.~A.~Rovenchak,
J.~Phys.~Stud. {\bf 4}, 16 (2000).\label{JPS2000}

\bibitem{Aziz79}
R.~A.~Aziz, V.~P.~S.~Nain, J.~S.~Carley, W.~L.~Taylor,
G.~T.~McConville, J.~Chem. Phys. {\bf 70}, 4430 (1979).

\bibitem{Aziz91}R.~A.~Aziz, M.~J.~Slaman,
J.~Chem. Phys. {\bf 94}, 8047 (1991).

\bibitem{Aziz92}R.~A.~Aziz, M.~J.~Slaman, A.~Koide, A.~R.~Allnatt,
W.~J.~Meath,
Mol. Phys. {\bf 77}, 321 (1992).

\bibitem{Aziz97}
A.~R.~Janzen and R.~A.~Aziz, J.~Chem. Phys. {\bf 107}, 914
(1997).\label{Aziz97}

\bibitem{Anderson93}J.~A.~Anderson, C.~A.~Traynor, B.~M.~Boghosian,
J.~Chem.~Phys. {\bf 99}, 345 (1993).\label{Anderson93}

\bibitem{Tang95}K.~T.~Tang, J.~P.~Toennies, C.~L.~Yiu,
Phys.~Rev.~B {\bf 74}, 1546 (1995).

\bibitem{Boronat94}J.~Boronat, J.~Casulleras,
Phys. Rev. B {\bf 49}, 8920 (1994).\label{Boronat94}

\bibitem{F2exp} Svensson~E.~C., Sears~V.~F., Woods~A.~D.~B., Martel~P.
Phys.~Rev.~B, {\bf 21}, 8 (1980). ???

\bibitem{ExpData} R. De Bruyn Ouboter, C.N. Yang, Physica B {\bf 44}, 127
(1987).\label{ExpData}

\bibitem{TMF2} I.~A.~Vakarchuk, Teor.~Mat.~Fiz. {\bf 80},
439 (1989); {\bf 82}, 438 (1990).

\bibitem{TMFS0} I.~A.~Vakarchuk, I.~R.~Yukhnovsky, Teor.~Mat.~Fiz. {\bf 40},
100 (1979).

\bibitem{Balescu}R.~Balescu, {\it Equilibrium and nonequilibrium statistical
mechanics} (New York: Wiley, 1975).

\bibitem{JPS1996} I.~O.~Vakarchuk, J.~Phys.~Stud. {\bf 1}, 25 (1996).

\bibitem{McCarthy} R.~D.~McCarthy, Nat. Bur. Stand. (U.S.), Tech.
Note 1029, (1980).

\bibitem{NIST} V.~D.~Arp, R.~D.~McCarty, D.~G.~Friend,
Natl. Inst. Stand. Technol. Tech. Note 1334 (revised) (1998).\label{NIST}

\end{thebibliography}
\end{document}